\documentstyle[12pt]{article}
\begin{document}
\def\etaten{\eta_{10}}
\def\he{${\rm {}^4He\:\:}$}
\def\be{\begin{equation}}
\def\ee{\end{equation}}
\def\bc{\begin{center}}
\def\ec{\end{center}}
\def\beq{\begin{eqnarray}}
\def\eeq{\end{eqnarray}}
\def\beqd{\begin{eqnarray*}}
\def\eeqd{\end{eqnarray*}}
\def\nin{\noindent}
\def\lra{$\leftrightarrow$ }
\def\eset{$\not\!\!\:0$ }
\def\bull{$\bullet$}
\def\reaceight{$^3$He$(\alpha , \gamma )^7$Be}
\def\reacnine{$^3$H$(\alpha , \gamma )^7$Li}
\def\dhe{D + $^3$He }
\def\dheh{(D + $^3$He)/H }
\def\omb{\Omega_B}
\pagestyle{empty}
\rightline{{\bf CWRU-P6-94}}
\rightline{APRIL 1994}
\baselineskip=16pt
\vskip 2.5in
\begin{center}
\bf\large {RECENT DEUTERIUM OBSERVATIONS AND BIG BANG
NUCLEOSYNTHESIS: A NEW PARADIGM?} 
\end{center}
\vskip0.2in
\begin{center}
Lawrence M. Krauss\footnote{Also Dept of Astronomy} and Peter J. Kernan
\vskip .1in
 {\small\it Department of Physics\\
Case Western Reserve University\\ 10900 Euclid Ave., Cleveland, OH
44106-7079}
\vskip 0.4in
\end{center}

\newpage
\pagestyle{plain}
\baselineskip=21pt

{ \it
A new observation of D in a primordial gas cloud, 
made using the high resolution spectrograph at the Keck telescope,
indicates an abundance $ D/H =(1.9-2.5) \times 10^{-4}$
\cite{SCHR}.
  Since
deuterium is destroyed by stars, and the predicted Big Bang
Nucleosynthesis (BBN) abundance falls monotonically with increasing
baryon density, deuterium places a reliable upper
limit on the baryon density of the universe.  Because the new measurement
is substationally larger than previous, galactic estimates, it
would force a reassessment of BBN predictions--- if it is confirmed. 
Using a new BBN Monte Carlo code and analysis technique \cite{KK} we
derive constraints implied by a lower limit of $D/H =1.9 \times
10^{-4}$.  We find 
$\Omega_B \le .0068h^{-2}$, which is definitively incompatible with
baryonic halo dark matter.  We also explore
implications of combining the D measurement with
other light element abundances.   $^7Li$ provides a lower bound,
$\Omega_B \ge .004h^{-2}$.  Also, the
initial $^4He$ mass fraction ($Y_p$) would have to be less
than
$23.5\%$, assuming 3 light neutrino species---in good agreement with
present best fits.  
Finally, observational upper limits of $Y_p \le
24
\%$ and $^7Li/H \le 2.3 \times 10^{-10}$ would allow the number of
 neutrinos to
be as big as 3.9.}

\vskip 0.2in
\baselineskip=22pt
  
Over the past decade the constraints derived from BBN analyses have
become almost uncomfortably stringent.  The vast improvement in the
measured neutron half-life \cite{helium} has dramatically reduced
uncertainties in the predicted $^4He$ abundance, for example. As a
result, the upper limit one might derive on $\Omega_B$ using $^4He$
abundance constraints has marched downward as estimates of the primordial
$^4He$ have decreased. Even the relatively recent consistent incorporation
of uncertainties in the analysis \cite{kraussrom,smithetal} has not
broadened the allowed region substantially.  Perhaps the most
significant observable in constraining theory however, has been the
combination \dhe.   Under the assumption that this combination
cannot have increased since it was initially created, the claimed upper
limit of $10^{-4}$ on the \dheh fraction places a lower limit
on $\Omega_B$ which is sufficiently large that the predicted $^4He$
abundance, which increases with $\Omega_B$, is large enough to exceed
some ``2 $\sigma$" estimates of the allowed upper limit on
$Y_p$.

Recently, we re-examined BBN constraints \cite{KK} and argued that they
were even stronger than previous estimates, due to several features:

\noindent
(a) The world average
for the neutron half-life ($\tau_N = 889
\pm 2.1 sec$ ) has an uncertainty which is almost twice as
small as that used in previous analyses.

\noindent
(b) Improvements in the BBN code, including a finer
integration of the nuclear abundances, and the
inclusion of ${M_N^{-1}}$ effects\cite{seck:accbbn}, result in a new
$\etaten$ -independent correction of $+.0031$ to 
$Y_p$. All corrections applied to date thus combine to produce a
net correction to $Y_p$ of $+.0006$.

\noindent
(c) Incorporating the fact that predicted elemental abundances are
correlated means that determining limits on cosmological parameters by
using the different abundances independently, as had been the case, is
not consistent.  In our own Monte-Carlo analysis, we
demonstrated that simultaneous use of constraints on both \he and 
\dhe could dramatically reduce the number of
models which were allowed at the ``$2 \sigma$" level.

The BBN constraints we derived on the basis of these results, using the
assumption that 
\he $<24 \%$ and \dheh $< 10^{-4}$, were extremely tight.  In
particular, a \he abundance of less than $23.8 \% $ was shown to be
inconsistent with these limits.  This value was uncomfortably close to
the claimed upper limit on \he.  In addition, we demonstrated that
greater than 3.04 effective light neutrino species in the relativistic
gas at the onset of BBN was ruled out.  This constraint is more severe
than it might first appear, as even a light sterile right handed neutrino
in a minimally extended standard model contributes more than this.

Of course, the weakest link in such an analysis is the assumed light
element abundance.  Measurements of \he, for example, are mostly
indirect, and subject to large systematic uncertainties.  As a result,
we argued that our refined BBN analysis might just as well suggest the
need for revision of the light element abundance estimates inferred from
observation instead of arguing for or against new nonstandard physics. 

The new claimed observation, by Songaila {\it et al}, of deuterium in a
primordial gas cloud, at a level \dheh $=1.9-2.5 \times 10^{-4}$ is
particularly exciting in this regard.  It has long been argued that any
present measurement of D provides a lower limit on its primordial
abundance because D is so fragile that it is easily destroyed in stars. 
Previously quoted abundance estimates of $10^{-5}$ led to a firm upper
bound on $ \eta_{10} <8$ (defined by $\Omega_{B} =.0036 h^{-2} (T/2.726)^3
\etaten \times 10^{10}$, where $T$ is the microwave background temperature
today, and $h$ defines the Hubble parameter $ H= 100h$ km/(Mpc sec)), 
which clearly established that
baryons could not close the universe.  The present observation, an order
of magnitude larger, is also a factor of two greater than the previous
upper limit on the combination \dhe.  As a result, the new
constraints one can derive on $\Omega_B$ will be much more severe.  In
addition, as we shall show, this changes the way we combine elemental
abundance limits to get constraints on cosmology and particle physics.

The system explored by
Songaila {\it et al} in principle provides a direct probe of unprocessed D.
 Nevertheless, the present measurement is not
yet definitive, and could be subject to systematic errors
\cite{SCHR}.  For example, an intervening gas cloud moving relative
to the first cloud at a small radial velocity could produce H absorption
lines which are shifted, and could mimic D absorption lines.  Moreover,
 while one might expect the D abundance in the primordial clouds would exceed
that in the galaxy, it is not clear how the large value obtained
can be reconciled with previous galactic estimates . 

Caveats notwithstanding, because of its
potential importance for altering BBN constraints it is
worthwhile to examine just how these constraints would change based on the
present measurement.   Having just updated our BBN
analysis, it seemed an opportune time to examine this situation, and we
present our results below. While the authors of
\cite{SCHR} point out that at worst their observation provides
a conservative upper bound on D, we argue here that it is appropriate,
from the point of view of BBN to use the lower limit of $1.9 \times
10^{-4}$ on the D fraction as a {\it lower limit} on the primordial
deuterium abundance.  If their result is correct, the
primordial D abundance must exceed this value, since it can only be
destroyed by processes since BBN.  If their result is in error, then it
is not clear that the old upper limit on \dhe should be abandoned just because
it is less conservative. 

We find that not only is the new upper limit on
$\Omega_B$ implied by the new D value is low
enough to convincingly rule out baryonic dark matter,
constraints on both $Y_p$ and $N_{\nu}$ are also significantly altered. 
Finally, $^7$Li now provides a lower bound on $\Omega_B$, which while 
somewhat
uncertain, is compatible with
$\Omega_B$ comparable to the
luminous baryon abundance today.

Our Monte Carlo analysis proceeds as follows: we allow input
uncertainties corresponding to measurement uncertainties in the BBN
nuclear reactions.   Each rate is determined using a Gaussian
distributed random multiplier centered on unity, with a $1-\sigma$ width
based on that quoted in \cite{smithetal}. For the 
rates with temperature dependent uncertainties the original uniformly
distributed random number is saved and mapped into a new gaussian 
distribution with
the appropriate width
at each time step.  For each value of $\eta_{10}$ we then ran 1000 BBN
models. ( see
\cite{KK} for details).

In figure 1 we display the predicted ranges for D, \he, and $^7$Li for
the range $ 1< \etaten <2$, along with the new D observational lower
limit, and the claimed upper limits on \he and $^7Li$, assuming 3 light
neutrino species.  For each value of $\etaten$ all 1000 model predictions
are shown, along with the median predictions, the
one-side
$2
\sigma$ upper (lower) limit for D ($^7Li$), and the
symmetric $2
\sigma$ range for each element.  The one-side limits occur
when less than 50/1000 models fell below(above) the limits,
while the symmetric limits encompass the central 950/1000 predictions.   The D
and
$^7$Li limits imply the allowed range $ 1.13 <\etaten < 1.87$,
which leads to the constraints on $\Omega_B$ quoted above. (If one were to put
a upper limit on prmordial D of 2.5 $\times 10^{-4}$, derived
from the new D observation alone, the lower limit on $\Omega_B$ would increase 
by 15 $\%$ from the more conservative limit quoted above using $^7$Li) 

In figure 2, we display the predicted \he vs D abundances for the upper
limit $\etaten = 1.87$, assuming three species of light neutrinos.  Here
the clear anti-correlation between these abundances is clear.  Utilizing
this result, we see that for the region above the allowed D lower limit,
the maximal value of \he is near 23.5$\%$.  This is a complete
reversal of the previous BBN limits, which put a lower bound on \he. 
What is perhaps more interesting is that this new upper bound is far
more consistent with the best fit estimates of $ 23 \pm 1 \% (2 \sigma)$
which are often applied to \he.

The role played by
varying the number of neutrinos is quite different than it was when one
combined an upper limit on D+$^3He$ with an upper limit on $^4$He, aside from a
relaxed upper bound on N$_{\nu}$ due to
the lower predicted
\he fraction in this range of
$\etaten$.  Before,
raising the number of neutrinos tightened BBN constraints, but now
it can actually relax them. In order to give the most conservative
limits on 
$\etaten$ we must allow the effective number of neutrinos to vary from
3, to account for possible new particles contributing to the radiation
gas during BBN at the fraction of a neutrino level.  

The point is that in the range of $\etaten$ of interest, increasing
$N_{\nu}$ monotonically increases all elemental abundances.  
At $\etaten=1.8$ an
increase of $N_{\nu}$ of 1 produces an increase of 5$\%$ in Y$_p$, 15$\%$ in
D/H and 23$\%$ in $^7$Li. 
This implies that an increase in the effective number of relativistic
species will increase the upper limit on $\Omega_B$ because the predicted
D abundance will increase.  However, at some point the upper bound on $\Omega_B$
from \he (or $^7$Li) will eclipse that from D.  By varying the number of
neutrinos, we find the maximum allowed value of $\etaten$ is 1.91, obtained for
3.4 effective neutrino species.

Increasing the number of neutrino species beyond 3.9 results in
violation of the observational limits for any range of $\etaten$, as
shown in figure 3, where we plot the number of models, out of 1000,
allowed by {\it simultaneously} employing the $^4$He and $^7$Li limits, as a
function of $\etaten$ for both 3.9 and 4.0 neutrinos. Clearly, any further
relaxation of either limit would weaken the bound, allowing 4.0 neutrinos. This
relaxed limit on the number of effective light neutrinos is quite significant
for model building, as it allows breathing room for new physics.  For example,
an extra scalar particle in thermal equilibrium at the BBN era would contribute
about .55 neutrinos.

As we have illustrated, the new D measurement could be quite exciting for
cosmology.  If confirmed, it will change the way we use primordial element
abundances to get cosmological and particle physics constraints, as we have
shown in the new confidence ranges we have derived.  In some sense, the new
cosmological constraints would be satisfying.  The allowed range of
$\Omega_B$ would overlap with the amount of visible matter in the Universe. 
What you see may be what you get, a result which has some attraction.
In addition, the  \he fraction can be much closer to what some people
have been claiming, and the constraint on plausible extra particles in the
radiation gas during BBN is relaxed.   At the very least, if the new result is
confirmed, the evidence for non-baryonic dark matter will have increased
dramatically, and the possible confrontation with MACHO searches will be
interesting to watch.

\newpage

\clearpage
\noindent {\bf Figure Captions}

\vskip 0.2in

\noindent Figure 1: BBN Monte Carlo predictions as a function of
$\etaten$ for \he, D, and $^7$Li.  Shown are symmetric $95\%$ confidence
limits on each elemental abundance, as well as one sided $95\%$ upper (lower)
limits for D ($^7$Li).  Also shown are observational limits,
including the new D lower limit, we use.  
\vskip 0.1 in

\noindent Figure 2: 1000 Monte Carlo BBN predictions for $Y_p$ and D/H
abundances. The vertical line corresponds to a lower limit on D/H of $1.9
\times 10^{-4}$

\vskip 0.1 in
\noindent Figure 3: Number of models (out of 1000 total models) 
which simulataneously satisfy the constraints
$Y_p \le 24 \%$ and $^7$Li/H $\le 2.3 \times 10^{-10}$, for 3.9 and 4.0
neutrinos.
 \clearpage

\end{document}